\documentclass[epj,twocolumn]{webofc}
\usepackage[varg]{txfonts}   
\woctitle{The Innermost Regions of Relativistic Jets and Their Magnetic Fields}

\begin{document}

\title{Relativistic jets in narrow-line Seyfert 1 galaxies. New discoveries and open questions}

\author{F. D'Ammando\inst{1}\fnsep\thanks{\email{dammando@ira.inaf.it}}, M. Orienti\inst{1}, J. Finke\inst{2}, J. Larsson\inst{3},
        M. Giroletti\inst{1}, on behalf of the {\em Fermi} Large Area Telescope Collaboration
}

\institute{INAF - Istituto di Radioastronomia, Via Gobetti 101, I-40129 Bologna, Italy
\and U.S. Naval Research Laboratory, Code 7653, 4555 Overlook Ave. SW, Washington, DC 20375-5352, USA
\and KTH, Department of Physics, and the Oskar Klein Centre, AlbaNova, SE-106 91 Stockholm, Sweden}

\abstract{Before the launch of the {\em Fermi} satellite only two classes of AGNs were known to produce relativistic jets and thus emit up to the
  $\gamma$-ray energy range: blazars and radio galaxies, both hosted in giant
  elliptical galaxies. The first four years of observations by the Large Area
  Telescope on board {\em Fermi} confirmed that these two are the most
  numerous classes of identified sources in the extragalactic $\gamma$-ray sky,
  but the discovery of $\gamma$-ray emission from 5 radio-loud narrow-line
  Seyfert 1 galaxies revealed the presence of a possible emerging third class
  of AGNs with relativistic jets. Considering that narrow-line Seyfert 1 galaxies seem to be typically hosted in
  spiral galaxy, this finding poses intriguing questions about the nature of
  these objects, the onset of production of relativistic jets, and
  the cosmological evolution of radio-loud AGN. Here, we discuss the
  radio-to-$\gamma$-rays properties of the $\gamma$-ray emitting narrow-line
  Seyfert 1 galaxies, also in comparison with the blazar scenario.} 
\maketitle
\section{Introduction}
\label{intro}
Active galactic nuclei (AGN) are the most luminous persistent sources of
high-energy radiation in the Universe. However, only a small percentage of AGNs are radio-loud, and
this characteristic is commonly ascribed to the presence of a relativistic
jet, roughly perpendicular to the accretion disk. Accretion of gas onto the
super-massive black hole (SMBH) is thought to power these collimated jets, even if the nature of the coupling
between the accretion disk and the jet is still among the outstanding open
questions in high-energy astrophysics \cite[e.g.][]{blandford01,
  meier03}. Certainly relativistic jets are the most extreme expression of the power than can be generated by a SMBH in the center of an AGN, with a large fraction of the power
emitted in $\gamma$ rays. 

Before the launch of the {\it Fermi} satellite only
two classes of AGNs were known to produce these structures and thus to emit
up to the $\gamma$-ray energy range: blazars and radio galaxies, both hosted
in giant elliptical galaxies \cite[]{blandford78}. The point spread
function and sensitivity of the Large Area Telescope (LAT) on board {\em Fermi} provided an unprecedented angular resolution at
high energies for localizing a large number of newly detected $\gamma$-ray
emitting sources. The first 4 years of observation by the {\it Fermi}-LAT  confirmed that the extragalactic $\gamma$-ray sky is dominated
by radio-loud AGNs, being mostly blazars and some radio galaxies \cite[]{nolan12,ackermann13}. However, the discovery by {\em
  Fermi}-LAT of variable $\gamma$-ray emission from a few radio-loud narrow-line Seyfert 1s (NLSy1s) revealed the presence of a
possible third class of AGNs with relativistic jets
\cite[]{abdo2009a,abdo2009b,abdo2009c}. On the contrary, no radio-quiet
Seyfert galaxies were detected in $\gamma$ rays until now
\cite[]{ackermann12b}. This finding poses intriguing questions about the
nature of radio-loud NLSy1s, the onset of production of relativistic jets, the
mechanisms of high-energy emission, and
the cosmological evolution of radio-loud AGN.

NLSy1 is a class of AGN identified by \cite{osterbrock85} and characterized by their optical properties: narrow permitted lines (FWHM
(H$\beta$) $<$ 2000 km s$^{-1}$) emitted from the broad line region (BLR),
[OIII]/H$\beta$ $<$ 3, and a bump due to Fe II \cite[see e.g.][for a review]{pogge00}. They also exhibit strong X-ray variability, steep X-ray spectra,
substantial soft X-ray excess and relatively high luminosity \citep{boller96,grupe10}. These characteristics point to systems with smaller masses of the central BH (10$^6$-10$^8$ M$_\odot$) and higher accretion rates
(close to or above the Eddington limit) with respect to blazars and radio
galaxies. NLSy1s are generally radio-quiet (radio-loudness\footnote{$R$ being defined as the ratio between the 1.4\,GHz
  and 4400\,\AA\, rest-frame flux densities} $R<$ 10), with only a small fraction of them
\cite[$<$ 7$\%$,][]{komossa06} classified as radio-loud, and objects with high
values of radio-loudness ($R>$ 100) are even more sparse ($\sim$2.5\%), while
generally $\sim$15$\%$
of quasars are radio-loud. In the past, several authors investigated the
peculiarities of the radio-loud NLSy1s with non-simultaneous radio-to-X-ray
data, suggesting similarities with the young stages of quasars or different
types of blazars \cite[e.g.][]{komossa06,yuan08,foschini09}. 

The firm confirmation of the existence of relativistic jets also in this
subclass of Seyfert
galaxies opened a large and unexplored research space for important discoveries for our knowledge of the
  AGNs, but brought with itself new challenging questions. What are the
  differences between this class of $\gamma$-ray emitting AGNs and blazars and radio
  galaxies? How do these objects
  fit into the blazar sequence? What is the origin of the radio-loudness? What are the parameters
  determining the jet formation? Is there a limiting BH mass above which
  objects are preferentially radio-loud? Five years after the announcement of
  the detection of the first NLSy1 in $\gamma$ rays by
{\it Fermi}-LAT, PMN J0948$+$0022 \cite[]{abdo2009a}, only some indications about the nature of these objects have been obtained.

\section{The $\gamma$-ray view of NLSy1}

Five radio-loud NLSy1 galaxies have been detected at high significance by {\em
  Fermi}-LAT so far: 1H 0323$+$342, SBS 0846$+$513, PMN J0948$+$0022, PKS
1502$+$036, and PKS 2004$-$447 \cite{nolan12, dammando12}, with a redshift
between 0.061 and 0.585. Here we analyze the first four years of $\gamma$-ray observations of these sources in two energy ranges: 0.1--1 GeV, and 1--100 GeV.
The LAT data reported in this paper were collected from 2008 August 4 (MJD
54682) to 2012 August 4 (MJD 56143). During this time the LAT instrument operated almost entirely in survey mode. 
The analysis was performed with the \texttt{ScienceTools} software package version v9r31p1. The LAT data
were extracted within a $10^{\circ}$ Region of Interest centred at the
location of the 5 NLSy1s. Only events belonging to the ``Source'' class were
used. The time intervals when the rocking angle of the LAT was greater than
52$^{\circ}$ were rejected.
In addition, a cut on the zenith angle ($< 100^{\circ}$) was also applied to reduce contamination from
the Earth limb $\gamma$ rays, which are produced by cosmic rays interacting with the upper atmosphere. 
The spectral analysis was performed with the instrument response functions
\texttt{P7SOURCE\_V6} using an unbinned maximum likelihood method implemented
in the Science tool \texttt{gtlike}. A Galactic diffuse emission model and
isotropic component, which is the sum of extragalactic and instrumental background were used to model the
background\footnote{http://fermi.gsfc.nasa.gov/ssc/data/access/lat/Background\\Models.html}. The
normalizations of both components in the background model were allowed to vary
freely during the spectral fitting.  

We evaluated the significance of the $\gamma$-ray signal from the sources by
means of the Test Statistics TS = 2$\Delta$log(likelihood) between models with
and without the source \cite[]{mattox96}. The source model used in
\texttt{gtlike} includes all the point sources from the 2FGL catalogue that
fall within $20^{\circ}$ from the target source. The spectra of these sources
were parametrized by power-law functions, $dN/dE \propto$
$(E/E_{0})^{-\Gamma}$, where $\Gamma$ is the photon index, or log-parabola,
$dN/dE \propto$ $(E/E_{0})^{-\alpha-\beta \, \log(E/E_0)}$, where $E_{0}$ is a
reference energy, $\alpha$ the spectral slope at the energy $E_{0}$, and the parameter $\beta$ measures the
curvature around the peak. 
A first maximum likelihood was performed to remove from the model the sources
having TS $<$ 25 and/or the predicted number of counts based on the fitted model $N_{pred} < 3 $. A second maximum likelihood was performed
on the updated source model. The fitting procedure has been performed with the
sources within 10$^{\circ}$ from the target source included with the
normalization factors and the photon indices left as free parameters. For the
sources located between 10$^{\circ}$ and 20$^{\circ}$ from our target we kept the
normalization and the photon index fixed to the values of the 2FGL
catalog. We used a power-law model for the five NLSy1s, except for the analysis
in the 0.1--1 GeV energy range of 1H 0323$+$342 and PMN J0948$+$0022, for
which we used a log-parabola as in the 2FGL
catalog \cite{nolan12}. The systematic uncertainty in the flux is energy dependent: it amounts to $10\%$ at 100 MeV, decreasing to
$5\%$ at 560 MeV, and increasing again to $10\%$ above 10 GeV \citep{ackermann12}.

The results of the LAT analysis over 2008 August--2012 August in the energy range 0.1--1 GeV and 1--100 GeV are summarized
in Table~\ref{LAT1} and Table~\ref{LAT2}, respectively. It is worth noting that the
detection significance of these NLSy1s is much higher in the 0.1--1 GeV energy
range with respect to the 1--100 GeV energy range, except for SBS
0846$+$513. That source is detected at high significance above 1 GeV with a
flat spectrum below 1 GeV, suggesting that during a strong flaring activity
SBS 0846$+$513 could be detected also at Very High Energy with the current Cherenkov 
Telescopes (MAGIC, VERITAS, H.E.S.S.) and the future Cherenkov Telescope Array
(CTA), as well as some flat spectrum radio quasars (FSRQs) at redshift comparable to that of this NLSy1 (i.e.~3C 279, 4C $+$21.35, and PKS
1510$-$089). On the contrary, only TS = 12 was obtained for 1H 0323$+$342
above 1 GeV, indicating that most of the $\gamma$-ray emission from this
source is usually produced at low $\gamma$-ray energies. In addition, 1H 0323$+$342 and PMN J0948$+$0022 showed a photon index higher
than 3 in the 1--100 GeV range, in agreement with a softening of the spectrum
due to a significant spectral curvature \cite[see 2FGL catalogue,][]{nolan12}.

As already reported in \cite{dammando13b}, the average apparent isotropic luminosity
of the 5 $\gamma$-rays NLSy1s estimated in the 0.1--100 GeV energy band spans between
10$^{44}$ erg s$^{-1}$ and 10$^{47}$ erg s$^{-1}$, a range of values typical of blazars. This could be an indication of
a small viewing angle with respect to the jet axis and thus a high beaming
factor for the $\gamma$-ray emission, similarly to blazars. In particular, SBS
0846$+$513 and PMN J0948$+$0022 showed $\gamma$-ray flaring activity combined
with a moderate spectral evolution \cite{dammando12,foschini11}, a behaviour already observed in bright FSRQs and low-synchrotron-peaked (LSP) BL Lacs detected by {\em Fermi}-LAT \cite{abdo10}. On the contrary, most of the radio galaxies detected by LAT have an apparent
isotropic $\gamma$-ray luminosity lower than 10$^{44}$ erg s$^{-1}$, suggesting a smaller beaming factor and possibly a different structure of the
jet \citep{abdo10b}. Several strong $\gamma$-ray flares were observed
from SBS 0846$+$513 and PMN J0948$+$0022, reaching at the peak an apparent isotropic $\gamma$-ray
luminosity of $\sim$10$^{48}$ erg s$^{-1}$, comparable to that of the bright
FSRQs \citep{foschini11,dammando12,dammando13e}. Variability and spectral properties of
these two NLSy1s in $\gamma$ rays indicate a blazar-like behaviour. Recently,
an intense $\gamma$-ray flaring activity was observed by
{\em Fermi}--LAT also from 1H 0323$+$342 \cite{carpenter13}. This is another
indication that radio-loud NLSy1s are able to host relativistic jets as
powerful as those in blazars. 

\begin{table*}
\caption{Results of the {\em Fermi}--LAT analysis of the NLSy1s in the 0.1--1
  GeV energy range.}
\begin{center}
\begin{tabular}{cccccc}
\hline \hline
\multicolumn{1}{c}{Source} &
\multicolumn{1}{c}{Redshift} &
\multicolumn{1}{c}{Flux (0.1--1 GeV)} &
\multicolumn{1}{c}{Photon index} &
\multicolumn{1}{c}{Curvature} &
\multicolumn{1}{c}{TS}  \\
\multicolumn{1}{c}{} &
\multicolumn{1}{c}{z} &
\multicolumn{1}{c}{$\times$10$^{-8}$ ph cm$^{-2}$ s$^{-1}$} &
\multicolumn{1}{c}{$\Gamma$/$\alpha$} &
\multicolumn{1}{c}{$\beta$}
\\
\hline
1H 0323$+$342 & 0.061    & 3.77$\pm$0.80   & 2.77$\pm$0.16  & 0.34$\pm$0.25  & 155 \\
SBS 0846$+$513 & 0.5835  & 2.10$\pm$0.23   & 2.04$\pm$0.13  & -              & 248 \\
PMN J0948$+$0022 & 0.585 & 11.25$\pm$0.48  & 2.50$\pm$0.05  &  0.26$\pm$0.09 & 1763 \\
PKS 1502$+$036 & 0.409   & 4.14$\pm$0.45   &  2.60$\pm$0.13 & -              & 220 \\
PKS 2004$-$447 & 0.240   & 1.23$\pm$0.38   & 2.08$\pm$0.25  & -              & 56 \\
\hline
\hline
\end{tabular}
\end{center}
\label{LAT1}
\end{table*}

\begin{table*}
\caption{Results of the {\em Fermi}--LAT analysis of the NLSy1s in the 1--100 GeV energy range.}
\begin{center}
\begin{tabular}{cccc}
\hline \hline
\multicolumn{1}{c}{Source} &
\multicolumn{1}{c}{Flux (1--100 GeV)} &
\multicolumn{1}{c}{Photon index} &
\multicolumn{1}{c}{TS}  \\
\multicolumn{1}{c}{} &
\multicolumn{1}{c}{$\times$10$^{-9}$ ph cm$^{-2}$ s$^{-1}$} &
\multicolumn{1}{c}{$\Gamma$}
\\
\hline
1H 0323$+$342 & 0.41$\pm$0.14   & 3.26$\pm$0.19  & 12 \\
SBS 0846$+$513 & 2.02$\pm$0.16   & 2.54$\pm$0.09  & 447 \\
PMN J0948$+$0022 & 2.71$\pm$0.23   & 3.46$\pm$0.04  & 410 \\
PKS 1502$+$036 & 0.10$\pm$0.01   & 2.65$\pm$0.18 & 96 \\
PKS 2004$-$447 & 0.42$\pm$0.12   & 2.63$\pm$0.32  & 22 \\
\hline
\hline
\end{tabular}
\end{center}
\label{LAT2}
\end{table*}

\section{X-ray observations}

The X-ray spectra of NLSy1 are usually characterized by a steep photon index
($\Gamma_{\rm X}$ $>$ 2) \cite{grupe10}. On the contrary, a relatively hard X-ray spectrum was detected in the {\em Swift}/XRT observations of SBS
0846$+$513 \cite{dammando12,dammando13b}, PMN J0948$+$0022 \cite{foschini11},
1H 0323$+$342 \cite{dammando13c}, and PKS 1502$+$036 \citep{dammando13a}. This
suggests a significant contribution of inverse Compton radiation from a
relativistic jet, similar to what is found for FSRQs.

The high quality XMM-{\em Newton} observation of PMN J0948$+$0022 performed in
2011 May allowed us to study in detail its X-ray spectrum, as reported in
detail in \cite{dammando13d}. The spectral
modelling of the XMM data of PMN J0948$+$0022 shows that emission from the jet most
likely dominates the spectrum above $\sim$2~keV, while a soft X-ray excess is
evident in the low-energy part of the X-ray spectrum. The origin of the soft X-ray excess is still an open issue both in
radio-quiet and radio-loud AGN \citep{gierlinski04,crummy06}. Such a Seyfert component is a typical feature in the X-ray spectra of
radio-quiet NLSy1s, but quite unusual in jet-dominated AGNs, even if not
unique \cite[e.g.~the FSRQ PKS 1510$-$089,][]{kataoka08}. It was not
possible to distinguish between different models for the soft X-ray emission of PMN J0948$+$0022 on a statistical basis. Models
where the soft emission is partly produced by blurred reflection, or
Comptonisation of the thermal disc emission, or simply a steep power-law, all
provide good fits to the data. A multicolor thermal disc emission also gives a comparable fit,
but a too high temperature (kT = 0.18 keV) is necessary, which is incompatible with a
standard Shakura \& Sunyaev accretion disc \cite{dammando13d}. A similar soft X-ray excess was observed also in the
XMM-{\em Newton} observation of the other $\gamma$-ray NLSy1 PKS 2004$-$447 \cite{gallo06}.
 
\section{Radio properties}

Only a small fraction of NLSy1s are known to be radio-loud. For these sources,
flat radio spectrum and flux density variability suggested that several of
them could host relativistic jets \citep{zhou03, doi06, yuan08}. A core-jet
structure on pc scale was observed for SBS 0846$+$513 \cite{dammando12}, PKS
1502$+$036 \cite{dammando13b}, and PMN J0948$+$0022
\cite{giroletti11,dammando13d}, although the jet structure in the last two
sources is significantly fainter than that observed in SBS 0846$+$513. Based
on VLBA data at 1.4 GHz the interpretation of the radio properties of PKS 2004$-$447 is more uncertain, with a bright and compact component at the
easternmost edge of the source that may be either the core of a core-jet blazar or
a bright hotspot of an asymmetric young radio source \cite{orienti12}. The
possibility that PKS 2004$-$447 could be a compact steep-spectrum source was
suggested also by \cite{gallo06}.

The analysis of the 6-epoch data set of SBS 0846$+$513 collected by the Monitoring Of
Jets in Active galactic nuclei with VLBA Experiments (MOJAVE)
programme\footnote{The MOJAVE data archive is maintained at
  http://www.physics.purdue.edu/MOJAVE} during 2011–2013 indicates that a superluminal jet component is moving away from the
core with an apparent angular velocity of (0.27$\pm$0.02) mas
yr$^{-1}$, corresponding to (9.3$\pm$0.6)$c$ \cite{dammando13a}. This apparent
superluminal velocity strongly suggests the presence of boosting effects for
the jet of SBS 0846$+$513. On the contrary, VLBA observations did not detect apparent superluminal motion at 15 GHz for PKS
1502+036 during 2002--2012, although the radio spectral variability and the
one-sided jet-like structure seem to require the presence of boosting effects
in a relativistic jet \cite{dammando13b}. 

\noindent In addition, strong variability was observed at 15 GHz during the monitoring of the Owens Valley Radio Observatory
40-m telescope of PMN J0948$+$0022 \cite{dammando13d}, PKS 1502$+$036 \cite{dammando13a}, and SBS
0846$+$513 \cite{dammando13e}. An inferred variability brightness temperature of
2.5$\times$10$^{13}$ K and 1.1$\times$10$^{14}$ K for PKS 1502$+$03 and SBS
0846$+$513, respectively. These values are much larger than the brightness temperature
derived for the Compton catastrophe \cite{kellermann69}, suggesting that the
radio emission of the jet is Doppler boosted. A complex connection between the
radio and $\gamma$-ray emission was observed for SBS 0846$+$513 and PMN
J0948$+$0022, as discussed in detail in \cite{dammando13e,orienti13}, and
\cite{dammando13d,foschini12}.

To understand if there are some peculiar characteristics in the 5 NLSy1s
detected by {\em Fermi}-LAT, we have investigated the radio properties of the first sample
of 23 radio-loud NLSy1s presented by \cite{yuan08}. We note that B3 1044$+$476, the object with the
highest radio-loudness, was not detected by {\em Fermi}-LAT in $\gamma$ rays so
far, while e.g. PMN J0948$+$0022 shows a high but not extreme radio-loudness. This indicates that the radio-loudness is not
necessarily a useful proxy for selecting the best candidates for $\gamma$-ray
detection with LAT. In the same way, a high apparent brightness temperature
of $\sim$10$^{13}$ K, comparable to that of SBS 0846$+$513 and PMN
J0948$+$0022, was observed for TXS 1546$+$353. This should be an indication of Doppler boosted emission
from a relativistic jet orientated close to our line-of-sight. However, no
$\gamma$-ray emission has been detected from this source. Certainly also the flux variability should be taken into account when searching for NLSy1s in $\gamma$
rays, but it is not the only factor playing a role. Indeed the discoveries of NLSy1s did not always occur during high $\gamma$-ray activity states (e.g. PKS 1502$+$036 and PKS 2004$-$447) \cite{abdo2009c,dammando13b}.

\section{Spectral energy distribution of $\gamma$-ray NLSy1}

The first spectral energy distributions (SEDs) collected for the four
$\gamma$-ray NLSy1s detected in the first year of {\em Fermi} operation showed
clear similarities with blazars: a double-humped shape with a first peak in the IR/optical band
due to synchrotron emission, a second peak in the MeV/GeV band likely due to inverse
Compton emission, and an accretion disk component in UV. The physical
parameters of these NLSy1s are blazar-like, and the jet power is in the average range of blazars \cite{abdo2009b}. 

The comparison of the SED of PMN J0948$+$0022 during the 2010 July flaring
activity with that of the FSRQ 3C 273 shows a more extreme Compton dominance in the NLSy1. The disagreement of the two SEDs may be due to the
difference in BH masses and Doppler factor of the two jets \citep{foschini11}. 

We also compared the SED of SBS 0846$+$513 during the flaring state in 2012
May with that of a quiescent state. The SEDs of the two different activity states,
modelled by an external Compton component of seed photons from a dust torus,
could be fitted by changing the electron distribution parameters as well as
the magnetic field \cite{dammando13e}, consistent with the modeling of different activity states of PKS\,0208$-$512 \cite{chatterjee13}. A significant shift of the synchrotron peak to higher frequencies was observed during the 2012 May flaring episode, similar to
  FSRQs \cite[e.g.~PKS\,1510$-$089;][]{dammando11}. Contrary to what is observed in PMN J0948$+$0022, no significant evidence of thermal emission from the accretion
  disc has been observed in SBS 0846$+$513 \cite{dammando13e}.

\section{Radio-loudness, host galaxies, and jet formation}

The mechanism at work for producing a relativistic jet is not still clear. In particular the physical parameters that
drive the jet formation is under debate yet. One fundamental parameter could
be the BH mass, with only large masses allowing an
efficient jet formation \cite[see e.g.][]{sikora07}. Therefore one of the most
surprising fact related to the discovery of the NLSy1s in $\gamma$ rays
was the development of a relativistic jet in objects with a relatively small
BH mass of 10$^{7}$-10$^{8}$ M$_{\odot}$ \cite{yuan08}. However, it is worth nothing that the mass
estimation of these source has large
uncertainties. In particular, \cite{marconi08} suggested that BLR clouds are subjected
to radiation pressure from the absorption of ionizing photons, and applying a
correction for this effect on the virial BH masses estimates higher masses are obtained for
the NLSy1s, which are objects radiating close to their Eddington limit.
In the same way, \cite{decarli08} proposed that the BLR may have a disk-like
geometry oriented almost face-on, so that the Doppler shifted line velocity projected along the line-of-sight appears small. 
Recently, \cite{calderone13} modelling the optical/UV data of some
radio-loud NLSy1s with a Shakura \& Sunyaev disc spectrum have estimated BH masses higher than 10$^{8}$ M$_{\odot}$. In particular, they found a BH mass
of 10$^{9}$ M$_{\odot}$ and 2$\times$10$^{8}$ M$_{\odot}$ for PMN J0948$+$0022
and PKS 1502$+$036, respectively, in agreement with the typical BH mass of blazars. This
may solve the problem of the minimum BH mass predicted in different scenarios of
relativistic jet formation and development, but introduces a new problem. If the
BH mass of these NLSy1s is 10$^{8}$-10$^{9}$ M$_{\odot}$, how is it possible to
have such a large BH mass in a class of AGNs usually hosted in spiral galaxies?

Unfortunately only very sparse observations of the host galaxy of radio-loud NLSy1s are available up to
now and the sample of objects studied by \cite{deo06} and \cite{zhou06} had z $<$ 0.03 and z $<$ 0.1, respectively, while four out five of
the NLSy1s detected in $\gamma$ rays have $z > 0.2$. Among the radio-loud NLSy1s detected by {\em Fermi}-LAT only for the closest one, 1H
0323$+$342, the host galaxy was clearly detected. {\em Hubble Space Telescope} (HST)
and Nordic Optical Telescope observations seem to reveal a one-armed galaxy
morphology or a circumnuclear ring, respectively, suggesting two possibilities: the spiral arms of the host
galaxy \cite[]{zhou07} or the residual of a galaxy merging \cite[]{anton08}. On
the other hand, no significant resolved structures have been observed by HST for SBS
0846$+$513 \cite[]{maoz93}, and no high-resolution observations are available for the
remaining $\gamma$-ray NLSy1s. Thus the possibility that the development of
relativistic jets in these objects could be due to strong merger activity, unusual in disk/spiral galaxies, cannot be ruled out. 

According to the ``modified spin paradigm'' proposed, another fundamental
parameter for the efficiency of a relativistic jet production should be the BH
spin, with SMBHs in elliptical galaxies having on average much larger spins than SMBHs in spiral galaxies. This is due
to the fact that the spiral galaxies are characterized by multiple accretion
events with random angular momentum orientation and small increments
of mass, while elliptical galaxies underwent at least one major merger with
large matter accretion triggering an efficient spin-up of the SMBHs. 
The accretion rate (thus the mass) and the spin of the BH seem to
be related to the host galaxy, leading to the hypothesis that relativistic jets
can develop only in elliptical galaxy \cite[e.g.][]{marscher09,bottcher02}. We
noted that BH masses of radio-loud NLSy1s are generally larger with respect to
the entire sample of NLSy1s \cite[M$_{\rm BH}$ $\approx$(2--10) $\times$10$^{7}$ M$_\odot$;][]{komossa06,yuan08}, even if still small if
compared to radio-loud quasars. The larger BH masses of radio-loud NLSy1s with
respect to radio-quiet NLSy1s could be related to prolonged accretion episodes that can spin-up the BHs. In
this context, the small fraction of radio-loud NLSy1s with respect to radio-loud quasars could be
an indication that not in all of the former the high-accretion regime lasted
long enough to spin-up the central BH \cite[]{sikora09}.

\section{Concluding remarks}

The presence of a relativistic jet in some radio-loud NLSy1 galaxies, first
suggested by their variable radio emission and flat radio spectra, is now
confirmed by the {\em Fermi}-LAT detection of five NLSy1s
in $\gamma$ rays. The flaring episodes observed in $\gamma$ rays from SBS
0846$+$513, PMN J0948$+$0022, and 1H 0323$+$342 are strong indications of a relativistic jet in these objects as
powerful as those of blazars. Variability and spectral properties in radio and
$\gamma$-ray bands, together with the SED modeling, indicate blazar-like behaviour.
These sources showed all the characteristics of the blazar phenomenon and they could be at the low tail of the blazar's BH
mass distribution, although it must be taken into account the possibility
that the masses of these radio-loud NLSy1s may be underestimated. 

Further multifrequency observations of the $\gamma$-ray emitting NLSy1s will be fundamental for
investigating in detail their characteristics over the entire electromagnetic
spectrum. The impact of the properties of the central engine in radio-loud
NLSy1s, which seem quite different from that in quasars and manifest in their
peculiar optical characteristics, on the $\gamma$-ray production mechanisms is
currently under debate. In addition, the detection of relativistic jets in a class of AGN usually hosted in spiral galaxies is very
intriguing, challenging the theoretical scenario of relativistic jet formation
proposed so far. 

Anyway, the lack of information about the host galaxy for four out five NLSy1s detected by LAT, together with the possible evidence the
residual of a galaxy merging for 1H 0323$+$342, left open the possibility that
at least some of the radio-loud NLSy1s are hosted in an early type elliptical/S0 galaxy. Further high-resolution 
observations of the host galaxy of $\gamma$-ray NLSy1s will be fundamental to obtain
important insights into the formation of relativistic jets.

\begin{acknowledgement}
The {\em Fermi} LAT Collaboration acknowledges generous ongoing
support from a number of agencies and institutes that have supported
both the development and the operation of the LAT as well as
scientific data analysis.  These include the National Aeronautics and
Space Administration and the Department of Energy in the United
States, the Commissariat \`a l'Energie Atomique and the Centre
National de la Recherche Scientifique / Institut National de Physique
Nucl\'eaire et de Physique des Particules in France, the Agenzia
Spaziale Italiana and the Istituto Nazionale di Fisica Nucleare in
Italy, the Ministry of Education, Culture, Sports, Science and
Technology (MEXT), High Energy Accelerator Research Organization (KEK)
and Japan Aerospace Exploration Agency (JAXA) in Japan, and the
K.~A.~Wallenberg Foundation, the Swedish Research Council and the
Swedish National Space Board in Sweden. Additional support for science
analysis during the operations phase is gratefully acknowledged from
the Istituto Nazionale di Astrofisica in Italy and the Centre National
d'\'Etudes Spatiales in France. FD, MO, MG acknowledge financial contribution from
grant PRIN-INAF-2011. FD thank also C. M. Raiteri, A. Doi, L. Stawarz,
D. Dallacasa, T. Hovatta and the OVRO Team, E. Angelakis, L. Fuhrmann and the
F-GAMMA Team, M. Lister and the MOJAVE Team, A. Drake and the CRTS Team,
A. L\"ahteenm\"aki, E. Lindfors and the Mets\"ahovi Team, for all the fruitful work
done together on this topic.
\end{acknowledgement}
 
\bigskip 

\end{document}